\documentclass [12pt]{article}
\def\be{\begin{equation}}
\def\ee{\end{equation}}
\def\bea{\begin{eqnarray}}
\def\eea{\end{eqnarray}}

\def\GeV{\mbox{GeV}}
\def\G{\mbox{G}}
\def\yr{\mbox{yr}}
\def\km{\mbox{km}}
\def\s{\mbox{s}}

\def\cm{\mbox{cm}}

\begin{document}
   

\begin{center}
{\LARGE \bf A study of diffusive shock acceleration as a process explaining observations of 1\, E0657-56 galaxy cluster}

\vspace*{1cm}
{G. Siemieniec-Ozi\c{e}b\l{}o}

\vspace*{1cm}
{\small \sl Astronomical Observatory, Jagellonian University}\\
{\small \sl Faculty of Physics, Astronomy and Applied Computer Science}\\
{\small \sl ul. Orla 171, 30--244 Krak\'ow, Poland}\\
{\small \sl (E-mail: grazyna@oa.uj.edu.pl)}

\end{center}
        


\vspace*{1cm}
\begin{abstract}
Chandra X-ray observations of the 1 E0657-56 galaxy cluster diffuse emission reveal the  existence of large scale cluster merger shock. We study the observed radio and X-ray data of this cluster as a possible result of the diffusive shock acceleration. This model can explain the observations within reasonable ranges of physical parameters for the shock and the intracluster medium. 
The expected nonthermal soft and hard X-ray fluxes are predicted.
\end{abstract}

\newpage
   
\section{Introduction}
Extended diffuse emissions which have recently been observed in clusters of galaxies provide an  evidence for the presence of nonthermal particles. The radio halo observations (Giovannini and Feretti, 2000) often show close similarity to the spatial distributions of X-ray thermal component present in these clusters (Giovannini et al., 1999). In a number of cases, nonthermal hard X-ray emission (HXR) (Fusco-Femiano et al. 2000) and an extreme ultraviolet emission (EUV) excess (Lieu et al. 1999) are also detected. The existence of the nonthermal X-ray component and its expected extension up to the $\gamma$-ray range is a challenge of consistent description of mechanisms generating the radiation in the intracluster medium (ICM). 
Extended synchrotron emission requires large scale magnetic fields and the respective electron cosmic ray (CR) component in the cluster. The expected contribution of energetic hadrons in radiative processes is still a matter of debate. It is not clear at present how energetic particles are accelerated in galaxy clusters. 

Determining the acceleration mechanism in galaxy clusters is essential for an understanding of the nonthermal phenomena in the ICM; both for the knowledge of the origin of radio relics and radio halos, through the ionized medium heating, up to the explanation of the observed high energy emission in X and $\gamma$-ray ranges.
In the literature many models were discussed concerning the nonthermal electrons and hadrons in clusters (Colafrancesco et al. 1998). The proposed acceleration processes are frequently associated with cluster merger phenomena (Sarazin 1999) and/or supergalactic accretion shocks (Ensslin et al. 1998). In another category of acceleration scenarios the particle scattering off magnetohydrodynamics turbulence in the ICM is the primary acceleration agent (Blasi 2000). With the extensive data base covering the radio and X-ray range, the best studied example for such processes is the Coma cluster. However, the observations of this object, and of many others, have not yet pin-pointed the dominant acceleration process.

Additional motivation for reconsidering existing ideas on particle acceleration is due to the new Chandra observations of the remarkably hot cluster, 1\,E0657-56 (Markevitch et al. 2002; P1), which reveal clear evidence of a merger shock. The cluster also posesses one of the most powerful radio halos (Liang et al. 2000; P2). Since it presents a classic example of a moderate shock produced during the merger, one may reconsider  the shock acceleration processes of electrons in this cluster.

In the present paper diffusive shock acceleration (DSA) is assumed to produce radio halo emission, and we will study the radio spectral properties and transport parameters of energetic electrons accelerated at the shock. The results of this study suggest that the DSA electrons are responsible for several major observational features in the cluster when reasonable transport coefficients are used. Finally, the implications of these relativistic electrons to nonthermal soft and hard X-ray emission due to inverse Compton (IC) processes are briefly discussed.

\section{Proposed model for acceleration at the merger shock of 1\, E0657-56}
\subsection{Radio halo emission}
Let us first list a few basic observational facts concerning the 1\,E0657-56 cluster. In the X-ray emission of this cluster the bullet-like gas cloud traversing the main cluster can be seen, after passing its centre $\sim 10^8$ yr ago (P1). It is a possible remnant of the merging subcluster. On-going merger observations allow determination of parameters of the shock seen as an edge brightness in the X-ray image. The independent indication of the shock presence is provided by the substantial temperature variations in the cluster. In particular, the temperature jump across the shock front is consistent with the Mach number, $M = 2.1$ and the estimated shock velocity $u_s \sim 3000 - 4000$ km $s^{-1}$, with the high mean gas temperature of 14 - 15 keV. The radio emission of the bullet substructure, the Western part of the  region marked `2' in (P2), including the bullet and the main cluster core, corresponds to the subvolume described above. It is characterized by spectral indices $\alpha_{4.9}^{1.3} \sim 1.2$ from $1.3$ to $4.9$ GHz and $\alpha_{8.8}^{2.4}\sim 1.3$ from $2.4$ to $8.8$~GHz, without any significant steepening. The radio power of this luminous halo is $P_{1.4} \sim 4.3\times10^{25}$ W /Hz at 1.4 GHz.

 Assembling the justifications for a shock-induced halo model, it is worth noticing that one cannot exclude the existence of a radio relic at the S-E peripheries of the 1\,E0657-56 cluster. Radio observations of the extended J\,06587-5558 radio source (Liang et al. 2001), placed near the initial merger site, i.e. 1.5 Mpc from the X-ray bow shock, allow that apart from the diffuse radio halo there may be a small-scale diffuse relic feature. Its location is seen at the S-E corner in Fig. 3 in (P1) and corresponds to the `discrete' source C in Fig. 4b in (P2). The polarization of the synchrotron emission observed from this peculiar source is exceptionally high and its spectral index $\alpha _{2.2}^{1.3} = 1$ steepens at higher frequencies to $\alpha _{8.8}^{4.8} = 1.5$. Whether it is possible to indicate a common origin of this hypothetical relic and the diffuse radio halo would be a question to answer after the nature of supposed relic has been confirmed.
 
 Observations of the ICM shock in this cluster suggest that the DSA model application may be relevant, leading possibly to formation of the synchrotron tail, a 2 Mpc structure up to an X-ray brightness edge. The presence of nonthermal particles responsible for the radio halo, and the simultaneous ICM gas heating during the merger event may be compatible with the shock dissipation scenario. In the standard DSA theory (e.g. Drury 1983) a few percent of the shock energy can be transmitted to cosmic rays while the remaining large portion is transferred to ICM heating.
 
In order to evaluate CR transport parameters the following assumptions are made:

1. The observed bullet-like structure in the X-ray data (P1), is a site of fresh electron injection. Its spectral behaviour represents the injected spectrum observed on time-scales shorter than cooling time for a particular energy. Thus it determines the spatial region $l_{diff}$, where diffusion is not affected by the cooling (see eq. 7, below). In this region one can expect to observe a nonthermal HXR excess due to the most energetic electrons.

2. Outside the above structure, the diffusing electrons change subsequently their spectral index (due to obvious steepening and the shock strength changes) until diffusion and losses produce the equilibrium spectrum. The HXR emission should disappear in this region.
 
 The efficiency of the DSA process depends mainly on the shock structure determined by its Mach number and  on the possibility of strong particle scattering in its vicinity. The physical properties of the merger shock initiated by cluster collision are estimated in the following, for the 1\,E0657-56, from the Rankine-Hugoniot relations. For the measured downstream and upstream plasma temperatures
the shock compression ratio $r = 2.22$ is obtained, corresponding to the Mach number $M \simeq 2$, consistent with the value $\approx 2.1$ predicted in (P1) and with a gas pressure jump (cf. Fig. 4 in (P1)- showing the gas pressure model) at the shock 
\begin{eqnarray}
{p_2 \over p_1} = {4r-1 \over 4-r} \approx 4.5 \ .
\end{eqnarray}
It proves that we do not have here a cold front but a real shock. The shocks like this, with moderate Mach numbers, allow for the most efficient ICM heating (Miniati 2002). Thus the extended radio emission, with morphology that resembles that in X-rays, becomes comprehensible. In particular, the radio emission seems to share the symmetry of the soft X-ray distribution.

A shock with the compression factor r can accelerate particles, producing the power law particle  momentum distribution, $n(p) \propto p ^{- \sigma}$, (see, e.g. Drury 1983), where the power law index
\begin{eqnarray}
\sigma = {r+2 \over r-1} \ .
\end{eqnarray}
The above compression factor implies a radio spectral index $\alpha = 1.25$ ($\sigma = 3.5$), in a  striking agreement with that measured at the central part of the radio halo, where $\alpha = (1.2 - 1.3)$ (P2). This fact encourages one to study the electron DSA model. To proceed, we assume, following (P2), that the magnetic energy density and the energy density of relativistic particles are close to equipartition. Then, a constant spectral index of $\approx 1.3$ above 0.01 GHz is taken, and a ratio of energy in cosmic ray  protons and electrons is assumed to be $k = 1$. The magnetic field strength is evaluated for the whole halo according to the standard formula (Miley 1980) as
\begin{eqnarray}
	B = 0.9 ~ \eta ^{2 \over 7}~ \mu \G \ ,
\end{eqnarray}
where $\eta$ is the magnetic field volume filling factor. In this formula a total flux density of 78 mJy at 1.3 GHz was used (P2). Below, $\eta \approx 1$ is used leading to $B \simeq 0.9 ~ \mu$G. For a strongly  filamentary magnetic field structure with $\eta = 0.01$ one obtains a somewhat larger field, $B \simeq 3.7 ~ \mu$G. 

Let us suppose that the energetic primary electrons generate the diffuse nonthermal radiation owing to acceleration occuring in on-going merger. The radio halo region corresponding to the area mentioned in point (1) i.e. from the cluster core to the shock front (cf. Fig.1a in (P1)) has a linear size  $\sim 0.6$ Mpc, (western extension of region `2' in (P2)). It is a fraction of the whole observed radio halo, believed to form the volume where DSA model offers a correct description. The shock propagating in the cluster distributes particles in a volume whose linear size corresponds to the distance  $\sim u_s t$, where $t \leq t_{life}$ ($t_{life}$ - the radiative life-time of relativistic electrons of the highest observed energy). In this volume, a young electron population with a spectrum identical to the injected one is assumed to produce the synchrotron emission with the observed index  1.2 - 1.3. The seed particles can be supplied at the merger shock.
  
There are two basic processes leading to energy loss for $\gamma \sim 10^3 - 10^4$  electrons: synchrotron and IC losses. Comparison of the acceleration time scale and the shorter of the loss scales allows evaluation of the maximum energy attainable by electrons. The energy losses for magnetic field calculated from eq. 3 (for simplicity let us take  $B \simeq 1 ~\mu$ G ) are dominated by the IC scattering of cosmic microwave background (CMB) photons. The respective time scale is given by
\begin{eqnarray}
	\tau_{IC} = {11.1 \times 10^8 \over (1+z)^4} \left( {E \over \GeV} \right)^{-1}= 3.88 \times 10^8 \left( {E \over \GeV} \right)^{-1} \yr,
\end{eqnarray}
where E is the electron energy and the cluster redshift is  z = 0.3. For a `moderate' shock with  $r \simeq 2$, the characteristic acceleration time scale is 
\begin{eqnarray}
	\tau_{acc} = {12 \over u_s^2} \: \kappa _{B}
\end{eqnarray}
where for highly turbulent conditions near the shock, the spatial diffusion coefficient $\kappa$ is close to the Bohm value,
\begin{eqnarray}
\kappa_{Bohm}={1 \over 3} r_g c=3.3 \times 10^{23} \left( {E \over \GeV} \right) \left( {B \over 10^{-7}\G} \right) ^{-1} \cm^2 \s^{-1} \nonumber
\end{eqnarray}
($r_g$ - a particle gyration radius). Thus we obtain
\begin{eqnarray}
	\tau_{acc} =12.54 \left( {E \over \GeV} \right) \left( {B \over 10^{-7}\G} \right) ^{-1} \left( {u_s \over 10^3~\km\s^{-1}} \right) ^{-2} \yr  .
\end{eqnarray}
The maximum energy condition for accelerated electrons, $\tau_{IC} = \tau_{acc}$, implies - for $B \sim 1 ~\mu$G and the  shock velocity $u_{s} \sim 4\times 10^3~ $km/s - the value of $E_{max} \approx 70$ TeV.
In the time $t_{life} = \tau_{IC}(E)$, the maximum distance swept by the shock gives for 3 GeV electrons radiating at 120 MHz -- 0.5 Mpc, and for 11 GeV electrons radiating at 1.4 GHz is 0.15 Mpc. It becomes clear that the explanation of synchrotron emission properties at 1.4 GeV is not trivial even if we concentrate exclusively on the mentioned western part of the cluster with the size of $\sim 0.6$ Mpc. 

The electron diffusion length in the limiting case of Bohm diffusion $\kappa = \kappa_{B}$ is for $t = t_{life}$
\begin{eqnarray}
	l_{diff} = (\kappa \ t_{life})^{1 \over 2} \ ,
\end{eqnarray} 
and gives $l_{diff}$ of only a few parsecs for 11 GeV electrons. However, a realistic description of particle transport in the region considered requires a detailed knowledge of spatial and energy dependence of the diffusion coefficient.

Owing to a lack of detailed information concerning the nature and geometry of the magnetized medium in the ICM, one must use simple considerations. In the merger shock vicinity the expected strong turbulence (Roettinger et al. 1999) implies that the Bohm approximation is valid. However, the particle mean free path far from the shock is expected to be much greater than in the shock  neighbourhood, leading to a more efficient diffusion. 
Thus one may ask a question about the value of the diffusion coefficient $\kappa$ which allows the shock accelerated electrons to fill the region of size  $L \sim 0.5 $ Mpc. Then, from eq. 7, for electrons between the energy 3 and 11 GeV,  the value of  $\kappa \sim$  $10^{32}~ \cm^2 s^{-1}$ is obtained. 
This value for $\kappa$ is much greater than $\kappa_B$. However, it should be noted, that the effective diffusion coefficient for cosmic rays in our Galaxy can reach $\sim 10^{28}~\cm^2/\s$, with an average cross-field particle transport. Thus the above estimate for the cluster medium, with the lower $B$ and the turbulence decaying far behind the shock, can be a realistic one.

Finally, let us briefly discuss the possible peripheral relic. If it is a real radio relic, the electrons contained there are expected to have an equilibrium spectrum $\sigma' = 3$ ($\alpha = 1$), corresponding to the injection spectrum $\sigma = 2$ created by the strong merger shock with the compression r = 4. The electron spectrum has to be already balanced by losses since the shock needed a long time to move from its original to the present position. At this  distance of the order of  1.5 Mpc, we expect its strength to be reduced to its present value,  r = 2. A decrease in compression and the following cooling result in a spectral index change   between the locations in question. For the electrons accelerated at the S-E cluster periphery one may  expect the observed spectrum steepening at higher frequencies, $\alpha_{8.8 }^{4.8 } = 1.5$, a  feature not seen at the present location of the shock, at the western ridge of region `2' in (P2). The steepening of the spectrum in the relic can be interpreted as the result 
of the magnetic field compression and enhancement by the merger, which
leads to much higher synchrotron losses. In such case, the compression 
would naturally explain a high polarization degree of this region.
 Since the shock dissipation leads to ICM gas heating and at the same time to decreasing the shock propagation velocity and the respective Mach number, a detailed {\it numerical} model describing the  process must consider its time dependence. Therefore, only an application of the time-dependent model will allow the entire radio halo to be treatable consistently.

\subsection{Nonthermal X-ray emission}

The same population of relativistic electrons that is responsible for the cluster radio halo generates nonthermal X-ray emission by IC scattering off cosmic microwave background photons. The emission may appear both at soft and hard X-ray energies as a power law excess. The hard X-ray emission  has been detected in several clusters (Rephaeli et al. 1999) with a luminosity of the order of $10^{43}$ erg$s^{-1}$.

To evaluate the luminosity of the nonthermal X-ray component in this cluster, the IC emissivity is calculated as (Baghi et al. 1998)
\begin{eqnarray}
	B_{\mu G} = 1.7 \: (1+z)^2 \left( {S_r \nu_r} \over {S_X \nu_X} \right) ^ {{2} \over {\sigma +1}}
\end{eqnarray}
for the ICM plasma parameters given above.
The radio and the X-ray flux densities, $S_r$ and $S_X$ are given at the observed frequencies $\nu_r$ [Hz] and $\nu_X$ [keV]. The IC radiation has the same spectral index $\alpha = 1.3$ ($\sigma = 3.5$) as the low frequency radio waves (assuming that there is no electron spectral break in the range corresponding to synchrotron emission within 10 MHz - 1.4 GHz); and the equipartition magnetic field strength, $B = 1~ \mu G$ is taken. 
For example, the electrons with $\gamma = 1500 - 4700$ scatter the CMB photons into the 2 -- 19 keV band,  $\nu_X \cong {4 \over 3}\nu_{CMB} \gamma^2 $. This results in a (cluster-frame) luminosity $L_{2-5~ keV} \sim 2 \times 10^{43}~$erg $s^{-1}$ and 
$L_{10-19 ~keV} \sim  10^{43}~$erg $s^{-1}$, respectively in the soft and the hard X-ray band. Thus the expected IC contribution to the X-ray intensity in the band 2-5 keV gives possibly a few percent of the total flux in this band. 

While the CMB photons should be pumped to HXR range, we expect that the same electrons, synchrotron radiating within the range 10 -- 1400 MHz, replenish the EUV band through the synchrotron self-Compton (SSC) process.  

The diffuse EUV emission is detected in excess of the thermal fluxes in the spectra of a few clusters (Bergh$\ddot{o}$fer et al. 2000). In the 1 E0657-56 cluster no nonthermal emission has yet  been detected - neither in the EUV range nor in any other.  (Although, as suggested by the X-ray spectral appearence in Fig. 2 in (P2), it might not be excluded for energies smaller than 0.4 keV or greater than 5 keV.) Since it would be too early to accept the detailed predictions of EUV excess, let us point out only that usually the EUV emission is attributed to IC scattering of CMB photons by nonthermal electrons with Lorentz factors of order of $\gamma \sim 400$ (Atoyan et al. 2000). We concentrate here on a $\gamma \geq 1000$ electron population (produced by on-going merger and not being the succesively stored older populations coming from past mergers), which supplies the luminous radio halo seen in radio frequencies. The lower energy part of the electron spectrum ($\gamma \leq 1000$) in a typical cluster presents a quite different distribution (see, for example Sarazin, 1999) than the power law high energy population. Therefore, an estimation of this emission, in e.g. the 80 - 400 eV band, produced by electrons with $\gamma \sim 300 - 700$ is not feasible through extrapolation of the spectrum to lower energies.

It must be mentioned that, apart from CMB, the dominant contributor to the cluster photon field, there are other radiation fields providing seed photons for IC scattering. However, as one can estimate for cluster starlight and radio photon field, their energy densities are always a few orders of magnitude smaller  than the CMB energy density. Thus both radiation fields give negligible contributions to HXR/gamma and EUV ranges.

\section{Summary}

A moderate shock seen in Chandra image of 1 E0657-56 galaxy cluster allows an application of a  diffusive shock acceleration model for the electron component of cosmic rays in this cluster. In the present paper an attempt has been made to consistently present the nonthermal phenomena  in this cluster in terms of DSA. The author would like to point out that the DSA model can explain observational spectral features of this cluster in a natural way. Further progress requires however an extension of the model to the nonstationary particle spectra as the consequence of shock braking in the cluster. In the case of the 1\,E0657-56 cluster, the evidence for a shock induced radio halo with possible support of {\it in situ} accelerated synchrotron emitting electrons is clearly seen. The parameters of particle acceleration evaluated from the observed shock characteristics are roughly consistent with the radio halo properties. Simple  estimates of IC fluxes in soft and hard X-ray bands are also provided. However, the decisive diagnostics of nonthermal ICM will become possible after inclusion of the hadronic interactions in the modelling.

\section{Acknowledgments}
The author is grateful to Michal Ostrowski for valuable discussions and for meticulous reading of the manuscript.

\end{document}